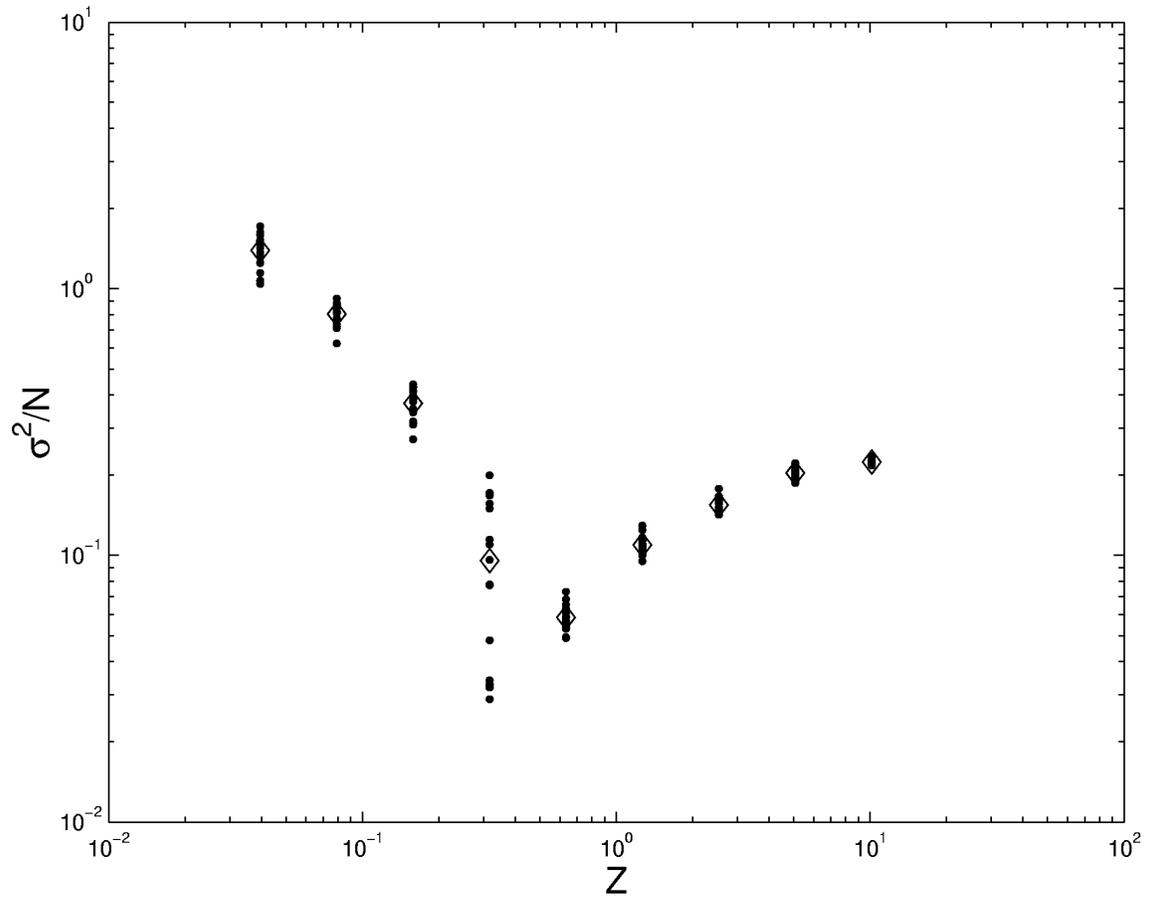

Figure 1

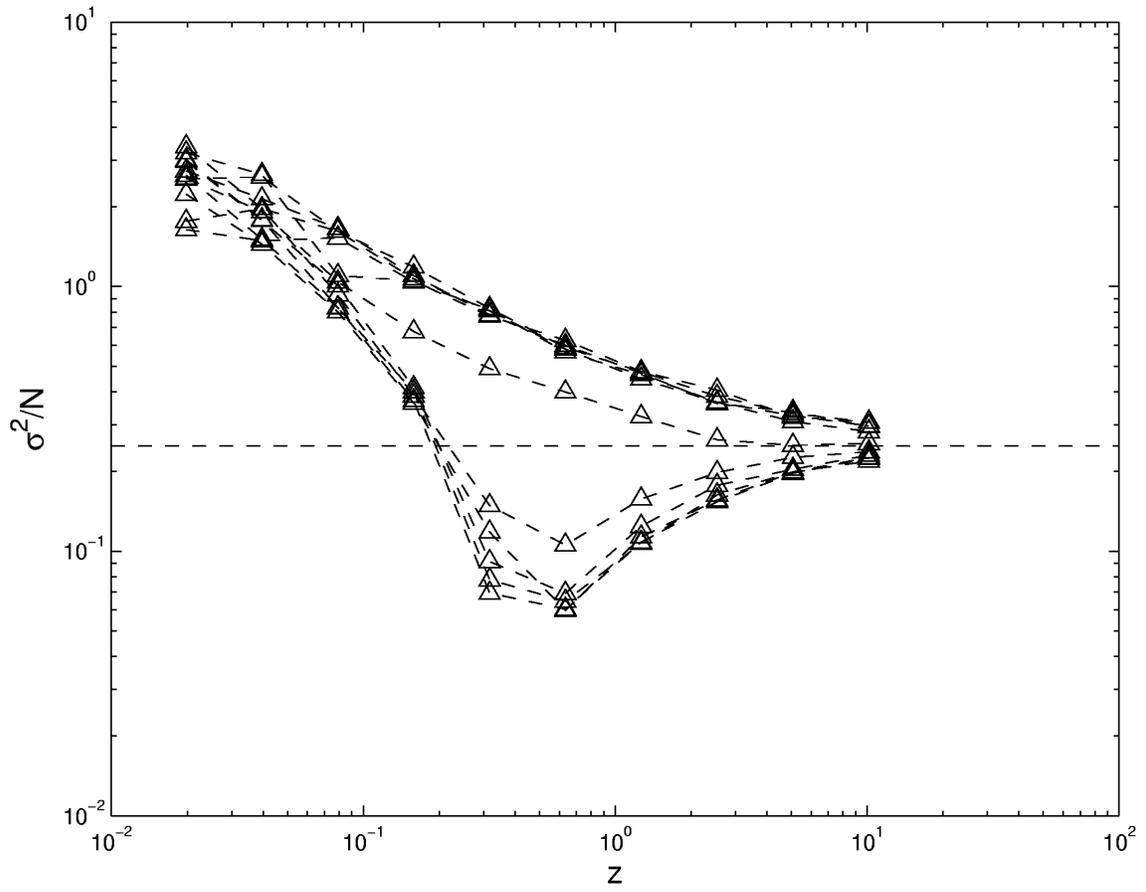

Figure 2

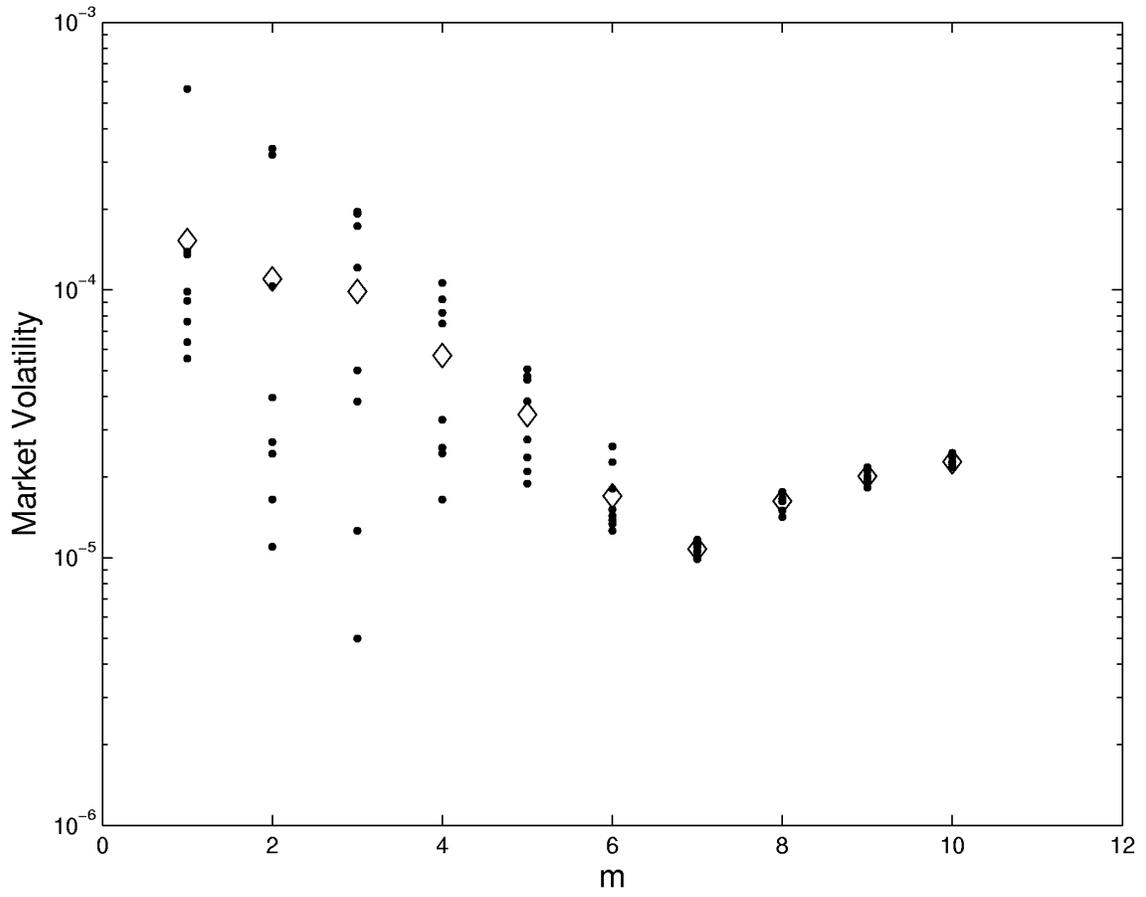

Figure 3

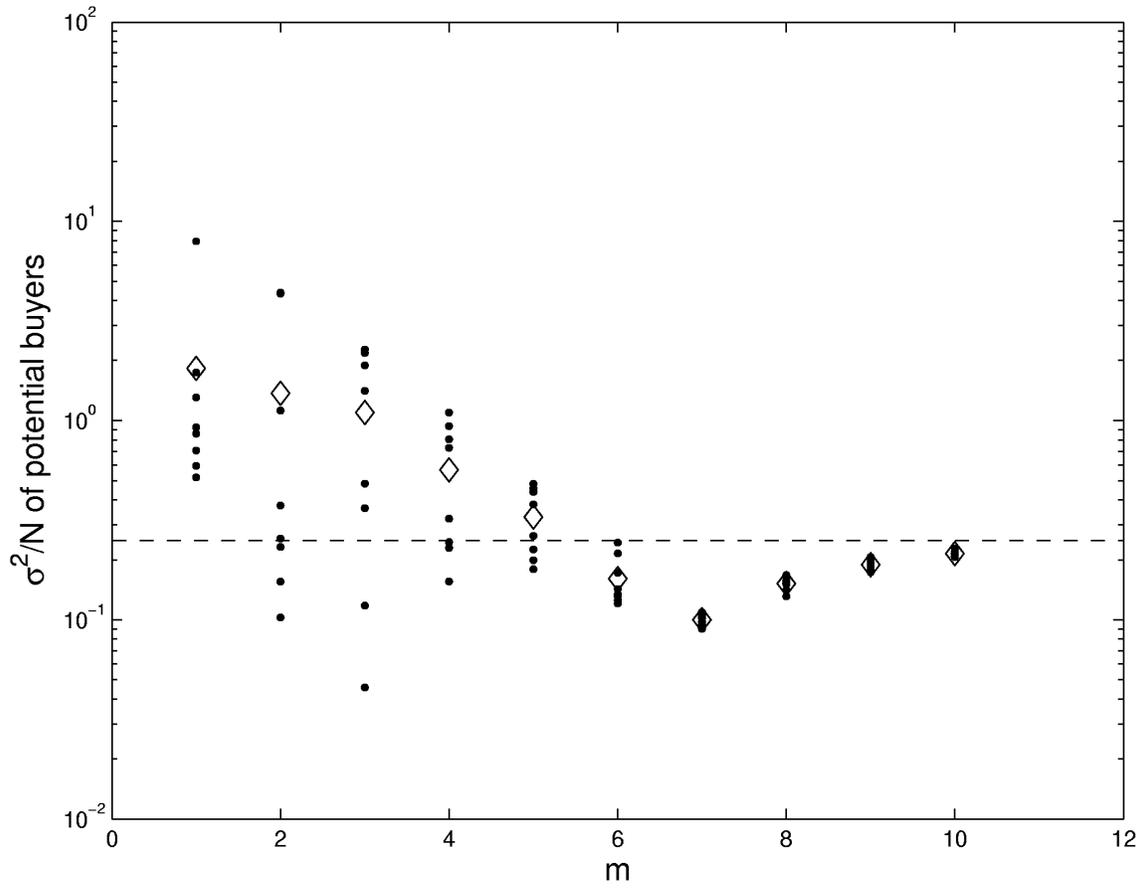

Figure 4

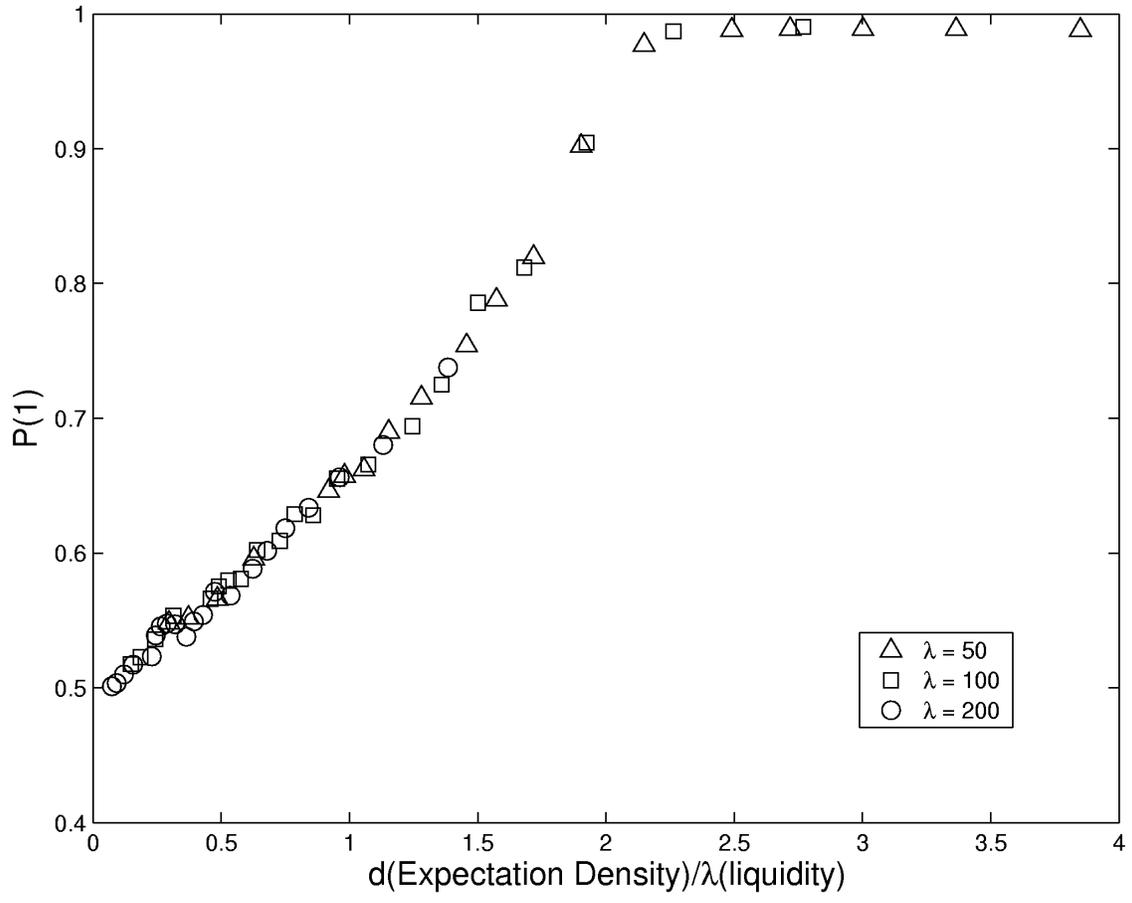

Figure 5

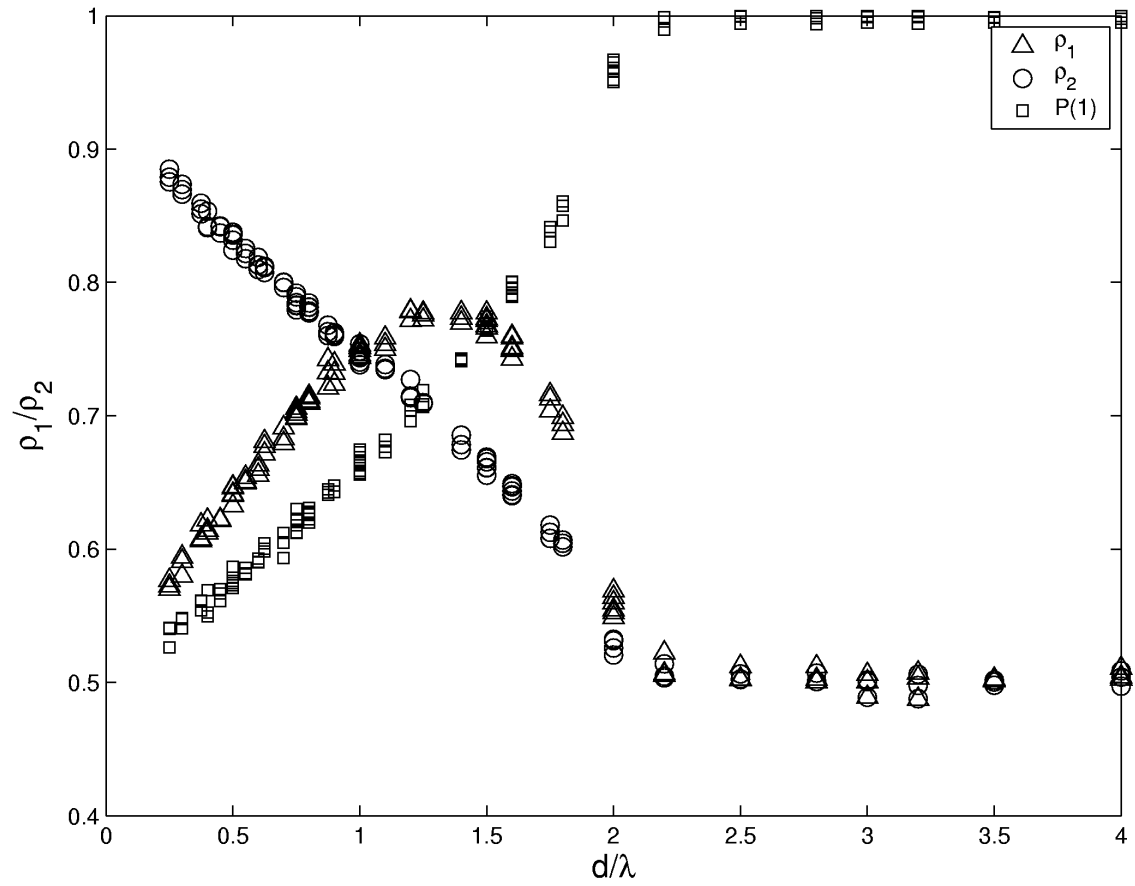

Figure 6

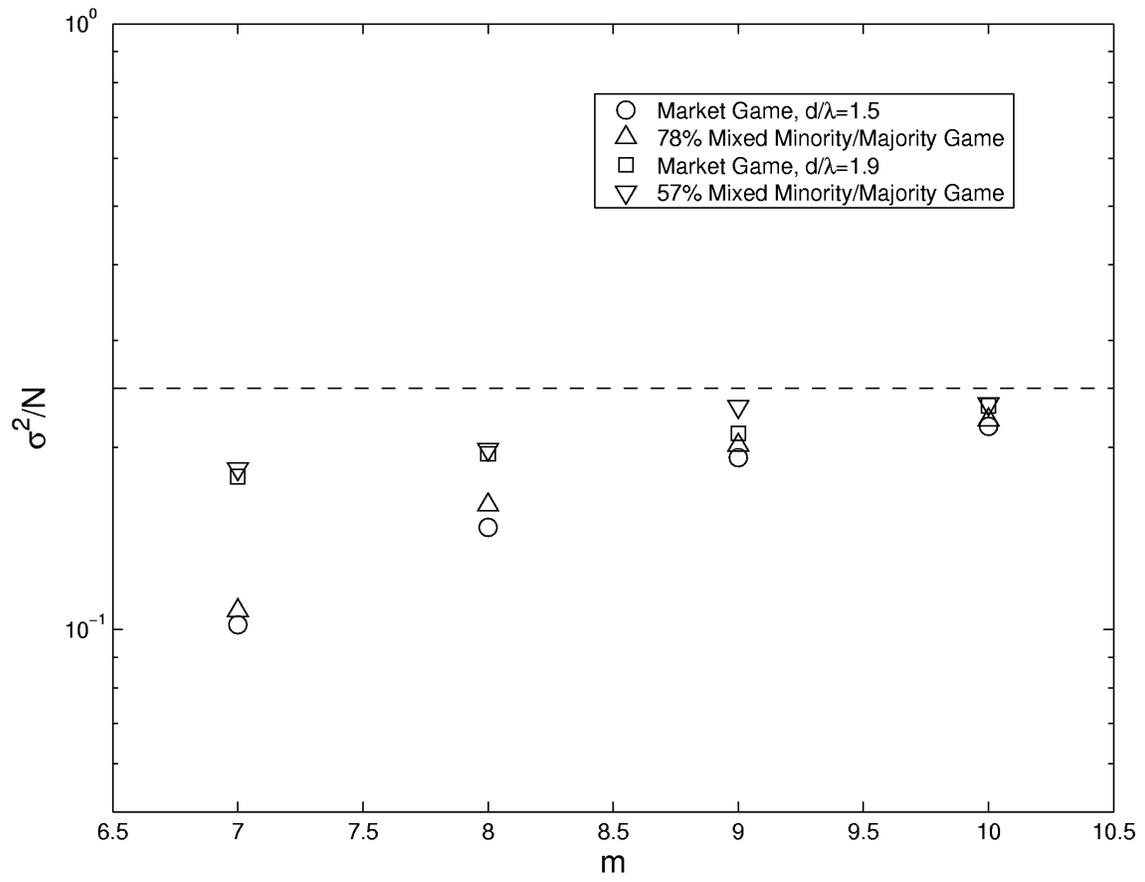

Figure 7

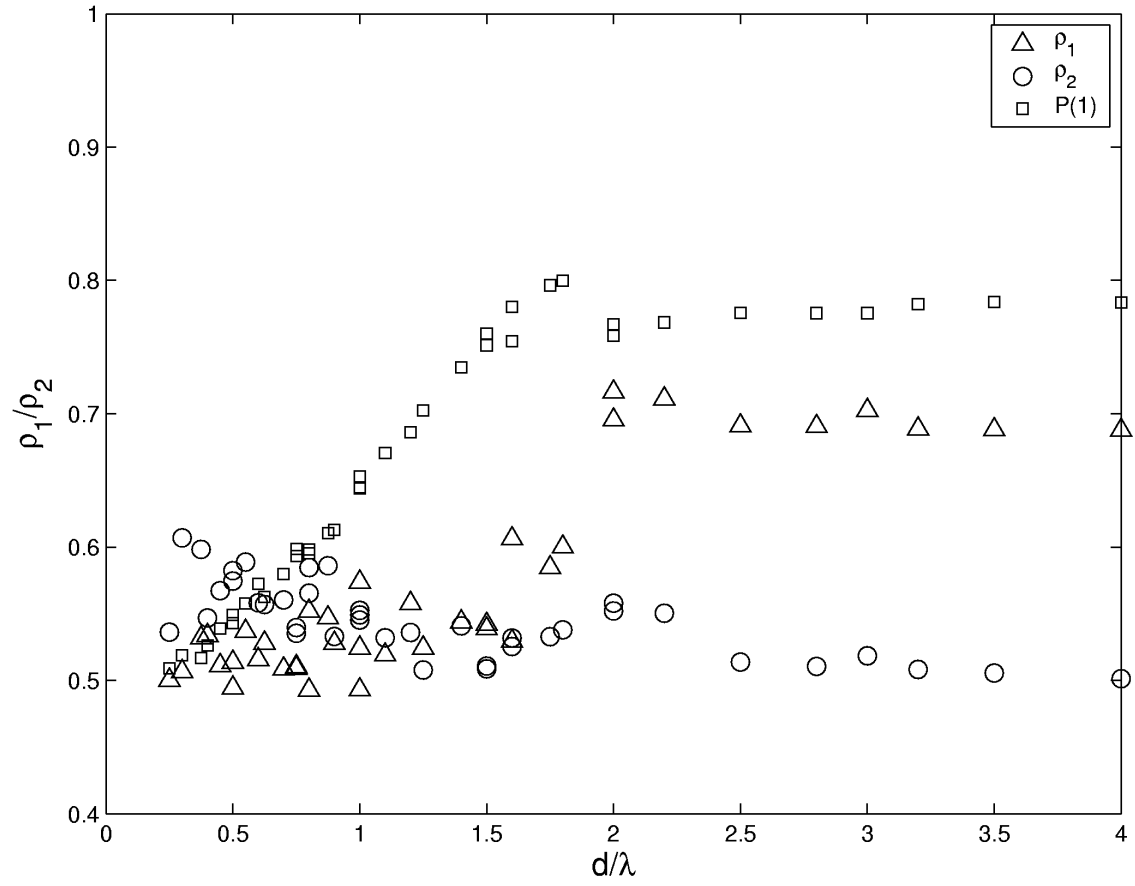

Figure 8

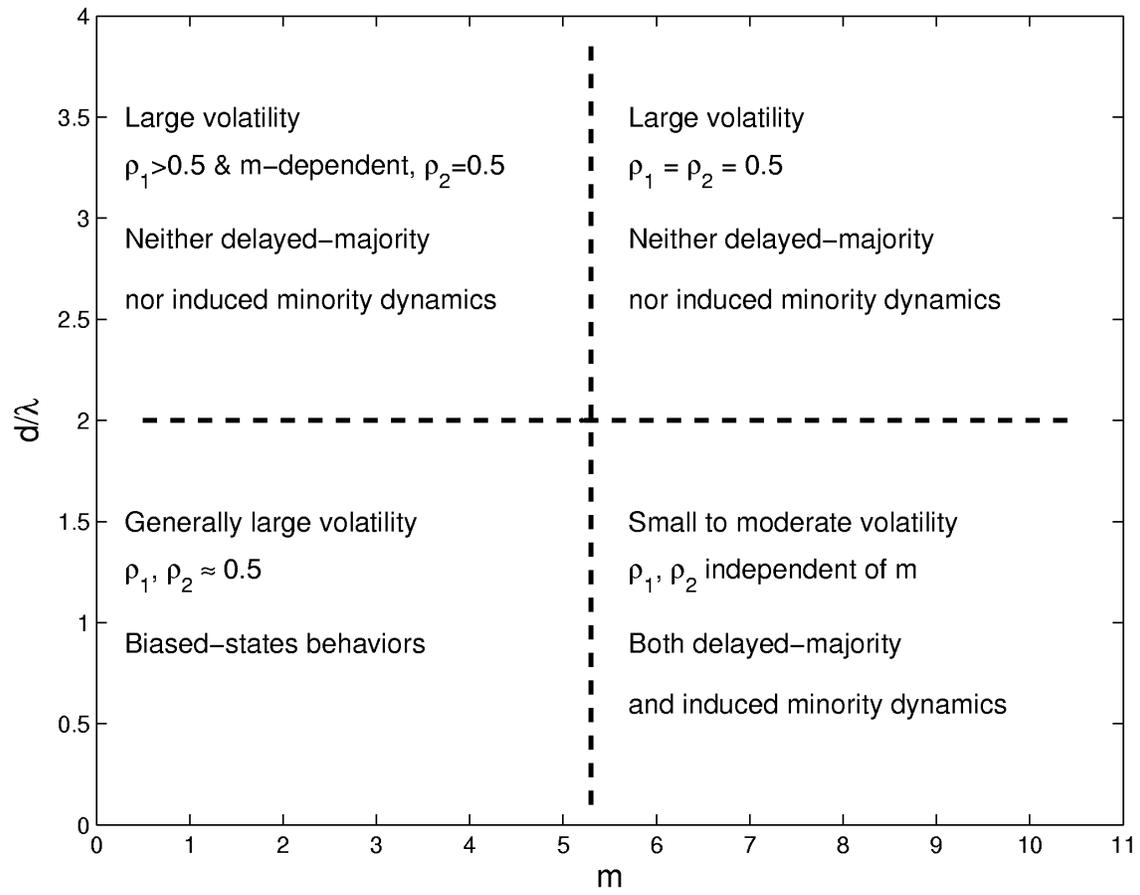

Figure 9

# Induced Minority Dynamics in a Stock Market Model

Yi Li and Robert Savit
Physics Department and Michigan Center for Theoretical Physics
University of Michigan
Ann Arbor, MI 48109

**ABSTRACT**

In this paper, we present a simple stock market model (the market game) which incorporates, as *ab initio* dynamics delayed majority dynamics, according to which agents (with heterogeneous strategies and price expectations) are rewarded if their actions at time t are the actions of the majority of agents at time t+1. We observe that for a range of parameter settings, minority dynamics are dynamically induced in this game, despite the fact that they are not introduced *ab initio*. Central to the emergence of minority dynamics is the introduction of the notion of price expectations for the agents. This leads to the possibility of an agent not participating in the market for some time steps. One consequence of the induced minority dynamics is an effective reduction in market volatility. We also discuss the phase structure and qualitative behavior of the market game for the entire parameter space.

12/19  v.6.4

## I.    Introduction

The behavior of agents seeking wealth or other rewards in the context of social and biological systems can be thought of as being composed of two types of paradigmatic actions: Innovation and conformity.  In the case of innovation, an agent will take an action that distinguishes himself from most other agents, i.e., one that places him in the minority.  Actions associated with conformity, on the other hand, are typically actions that place an agent in a group with most other agents, and can be thought of as "majority seeking".  Examples of the majority seeking include coalition politics, or generally conflicts in which large numbers of agents confer an advantage.  It is better to be a Democratic member of congress when the Democrats are the majority.  Other things being equal, it is also better to be a member of the larger of two battling armies.  In other circumstances, though, it is better to be in the minority.  In some examples of competition for limited resources agents that make minority choices will fare better.  For example, if there are two suppliers of parts, each with the same inventory, then those consumers that patronize the less popular of the suppliers will have a better chance of having their orders filled.  Similarly, driving to work along less crowded roads will often lead to shorter travel time and less psychological stress.

At first sight, it would appear that markets can also be thought of as embodying minority dynamics.  That is, if one invokes a simple relationship between price and supply vs. demand, then it would appear that participants in a market also benefit from being in the minority.  So, if there are more buyers than sellers, the price is high and the sellers (who benefit from a high price and are in the minority) will be rewarded.  On the other hand, if there are more sellers than buyers, the price will be low and the buyers who are in the minority will benefit.

A closer examination of the dynamics in markets, however, shows that the situation is not so simple.  In particular, depending on the way in which price is determined, the dynamics of a market may be more properly described as a variation of delayed majority dynamics.  That is, generically, an agent will benefit if his action at time t is adopted by a majority of agents at a later time.

In this paper we will show that, properly instantiated, models of a certain kind of market, which includes delayed majority dynamics can exhibit minority dynamics, even though minority behavior is not rewarded *per se*.  The ingredients necessary for this emergent minority dynamics are delayed majority dynamics and a mechanism that allows agents to drop out of trading if the commodity's current price violates the agent's beliefs about that commodity's underlying value.

The dynamical nature of the real financial markets and its relationship to the minority game have been the subject of some investigation over the past few years[1,2,3,4].  Among the early work in this area is an interesting project carried out by Marsili[1] that, superficially, resembles our work.  There are, however, fundamental differences. We think it is worthwhile to dedicate a short discussion at the beginning of section VI to the differences of the two approaches. The relationship of other work in this area to our investigations will be addressed at various, appropriate points in the paper.

In the next section we will first briefly review the simple minority and majority games, their definitions and fundamental behaviors. In section III we briefly discuss some mechanisms for price determination in different markets. This will motivate, in part, our introduction in section IV of the delayed majority game. In this section we will discuss the simple delayed majority dynamics and we will show that, in its simplest form its behavior is trivial. In section V we will introduce a mechanism based on prior (heterogeneous) beliefs about a commodity's worth which allows agents not to trade, and we will show that that mechanism, combined with the delayed majority dynamics induces minority game behavior in the system, even though minority dynamics is not explicitly included. The paper concludes with section VI which contains a summary and discussion of our results.

## II. Review of the Minority and Majority Games

Consider a game played by N (odd) agents. At each time step of the game, each of the N agents joins one of two groups, labeled 0 or 1. In the case of the minority game[5,6,7], each agent that is in the minority group at that time step is awarded a point, while each agent belonging to the majority group gets nothing. In the case of the majority game, each agent belonging to the majority group gets a point, while each agent belonging to the minority group gets nothing. An agent chooses which group to join at a given time step based on the prediction of a strategy. The strategy uses information from the historical record of which group was the minority [resp. majority] group as a function of time. A strategy of memory m is a table of 2 columns and $2^m$ rows. The left column contains all the $2^m$ possible combinations of m 0's and 1's, representing the list of which were the minority [resp. majority] groups for the past m time steps, while each entry in the right column is a 0 or a 1. To use this strategy, an agent observes which groups were the minority [resp. majority] groups during the immediately preceding m time steps, and finds that string of 0's and 1's in the left column of the table. The corresponding entry in the right column contains that strategy's determination of which group (0 or 1) will be the next minority [resp. majority] group, and thus which group the agent should join during the current time step.

In the simplest versions of the games, and in all of the games discussed in this paper, all strategies used by all the agents have the same value of m. At the beginning of the game each agent is randomly assigned s (>1) of the $2^{2^m}$ possible strategies, chosen with replacement. For his current play the agent chooses its strategy that would have had the best performance over the history of the game up to that time. In the case of the minority game, that strategy will be chosen that has most successfully predicted the minority group, while in the majority game, the strategy that has been most successful in predicting the majority group will be used. (Note that formally, this amounts merely to a reinterpretation of the nature of the group predicted by the strategy. However, this change results in markedly different dynamics in the two games.) Ties between strategies are decided by a coin toss.

The original version of the minority game was played endogenously—that is the time series of minority groups used by the agents to make their choices was the time series of the real minority groups generated endogenously by the play of the agents. Another version of the game uses exogenous information. That is, the information used by the agents to make their choices is generated exogenously, so that that particular time series does not necessarily reflect the actual outcome of the historical choices made by the agents. However, the rankings of an agent's strategies and the rewards to the agent are based on the *actual* minority groups. Although there are important differences, many of the most significant features of the endogenous and exogenous minority games are the same. In this paper we will primarily consider exogenous games. Where there are important qualitative differences with endogenous games we will note them. Finally, in all of the games described here each agent will have 2 strategies (s=2). Larger values of s generally change the results quantitatively, but not qualitatively.

The basic result of the standard minority game is shown in Fig. 1. Let $\sigma$ be the standard deviation of the number of agents in group 1. Fig. 1 shows $\sigma^2/N$ as a function of $z=2^m/N$ on a log-log scale for various N and m (with s=2). We see in this graph the, by now well known, phase transition at $z_c \approx 1/3$ separating two phases with qualitatively different behaviors. Also at $z_c$ we note that $\sigma^2/N$ is a minimum signifying that at this value of m the typical minority groups are larger than at other values of m. This means that when $z=z_c$ the system manifests the greatest degree of emergent coordination, resulting in the greatest wealth production, in the sense of generating the most number of points system-wide. Many variations of this game have been studied. The reader is referred to an excellent web site[8] for a guide to the literature.

The same structure can be used to play a simple majority game. The only difference with the minority game is that the strategies are interpreted to predict the next *majority* group. Agents are awarded a point for being in the majority group, and are awarded nothing if they are in the minority group. Strategies are ranked according to their success in predicting the majority groups. Different variations of the majority games have been exhaustively studied[9]. Here we restate only the major results that are pertinent to our present work. In the *endogenous* majority game, after a short transient period, the system reaches a stationary state in which the majority group is always 0 or always 1. In this case, the input signal is always '00…0' or '11…1'. In the *exogenous* majority game, however, the system reaches different stationary states in which each input signal corresponds to a fixed response that is not necessarily all the same. Agents, except those who have identical strategies, play with pure strategies, i.e., the strategy that can put the agent more frequently in a majority group. In this game, $\sigma^2/N$ is large and decreases with z. (Note, this description assumes that the probability distribution of strings in the exogenous majority game is stationary.) Also, there is no phase change as a function of N or m in either version of the majority game.

Another interesting variation is a version of a *mixed minority-majority* game. In this variation, the system rewards the minority group at some time steps, and rewards the majority group at other time steps. Which group to reward is decided in a random manner. Let us denote by *p* the fraction of times that majority group is rewarded. In fig. 2, we plot $\sigma^2/N$ against z for various *p*'s between 0 and 1. We observe that the hallmarks of the minority game (an apparent phase change and dip in $\sigma^2/N$) persist for games with small *p*. This behavior disappears at a *p* value slightly less than 0.5. For larger values of p, the behavior, a smoothly decreasing function of z, is qualitatively similar to that of the pure majority game. The transition from a dip structure to the smoothly decreasing curve occurs rapidly as p varies from about 0.4 to 0.5. This simple description of the behavior of the mixed minority/majority game is sufficient for our purposes. A detailed analysis of the mixed game will be presented elsewhere.[10]

It is worthwhile noting that other forms of mixed minority-majority game have been studied. In other versions of the mixed game[11], some of the agents are "minority players" (a.k.a. *fundamentalists*) who are rewarded if they are in the minority group, while other players are "majority players" (a.k.a. *trend-followers*), who are rewarded if they are in the majority group. In these games, the payoff scheme is consistent from timestep to timestep, so that minority (majority) players are always rewarded for being in the minority (majority). These schemes may be regarded as "mixed-population" schemes, while our mixing method presented is a "temporal-mixing" scheme. Interestingly, in both the mixed-population and temporal-mixing schemes, there is a critical mixing value around 0.5. That is, when the minority players exceed 50% of the population, or when the minority group is rewarded more than 50% of the time, the system exhibits the phase structure observed in the standard minority games.

### III. Markets and Price Determination

Although markets appear to exhibit some dynamics reminiscent of the minority game, the minority game itself is inadequate as a market model. In particular, the simple minority game *per se* does not account properly for the evaluation of portfolio value. To do that we need to introduce explicitly the idea of price and define more carefully the way in which price is determined. Our model of price and valuation follows.

Suppose we have a collection of agents trading some commodity. Let the commodity price at time t be p(t). Suppose also that, at time t, agent i holds $m_i(t)$ monetary units and $s_i(t)$ units of the commodity. Thus the value of the portfolio of agent i at time t is $v_i(t) = m_i(t) + s_i(t)p(t)$. Each trader can buy or sell one share at a time, and the excess demand is filled by the so-called *market makers*. We use a simple algorithm[12] to determine price movement:

$$\log p(t+1) = \log p(t) + D(t)/\lambda \qquad (1)$$

where $D(t) \equiv n_0(t) - n_1(t)$ is the excess demand at time t. $n_0(t)$ is the number of agents in group 0 (the buyers), $n_1(t)$ is the number of agents in group 1 (the sellers) at time t,), and $\lambda$ is the *liquidity coefficient* which determines the sensitivity of the price movement is to excess demand. In most of our games, we choose $\lambda$ in the range between 50.0 and 200.0. The effect of choosing different values of $\lambda$ will be discussed below.

To compute the changes in $v_i(t)$, we need to specify, in addition to the way in which prices move, the way in which the market clears. This will tell us the price that agents will need to pay for a unit of the commodity, and will thus relate changes in portfolio values to price changes. We consider here two options. First, the market can clear at the price that was set prior to the trade. In this case, at time t+1, buyers pay (and sellers receive) p(t) for a unit of commodity. This is an example of a *limit order*. On the other hand, for purposes of evaluating portfolios the commodity is valued at the current market price, i.e. at p(t+1). Under this scenario, it is easy to see that the net change in the value of a buyer's portfolio is (assuming that each agent's share is s before the trade)

$$\begin{aligned} \Delta v_i &\equiv v_i(t+1) - v_i(t) \\ &= -p(t) + (s_i(t)+1) \cdot p(t+1) - s_i(t) \cdot p(t) \\ &= (s_i(t)+1) \cdot (p(t+1) - p(t)) \end{aligned} \quad (2)$$

and a seller's portfolio changes in value by

$$\begin{aligned} \Delta v_i &\equiv v_i(t+1) - v_i(t) \\ &= p(t) + (s_i(t)-1) \cdot p(t+1) - s_i(t) \cdot p(t) \\ &= (s_i(t)-1) \cdot (p(t+1) - p(t)) \end{aligned} \quad (3)$$

If p(t+1)>p(t), the value of a buyer's portfolio increases, while if p(t+1)<p(t) the value of a seller's portfolio decreases. A price increase (decrease) results from a majority of buyers (sellers) at time t+1. Since buyers (sellers) benefit from a price increase (decrease) in the sense that the value their portfolios' increase, the dynamics of this kind of idealized market is that associated with a majority game.

The second method of market clearance we consider is that in which buyers pay the current market price p(t+1), which is determined by the mechanism of demand-supply relationship. This is the case of a *market order*. In this case, the net change in portfolio value for both sellers and buyers is

$$\Delta v_i \equiv v_i(t+1) - v_i(t) = s_i(t) \cdot (p(t+1) - p(t)). \quad (4)$$

Apparently, the market is neutral.

In this paper, we will focus on the case of the market order. It is clear that agents' trading histories will affect the value of their portfolios. For example, consider an agent who buys units of the commodity at time t. Suppose that at time t+1 there are more buyers than sellers. Then, according to (1) the price at time t+1 increases. Thus, even if the agent does not actively trade at time t+1, his portfolio will increase in value. Because this agent bought units of the commodity at time t, the value of his portfolio at time t+1 will increase more than if he had not bought the commodity at time t. Similarly, if p(t+1)<p(t) due to a surfeit of sellers at time t+1, the agent would suffer less loss had he sold some units of the commodity at time t. To summarize this simple analysis, the action of an agent at time t will be beneficial to him if that action is taken by the majority at time t+1. Buying (selling) at time t is advantageous over doing nothing or selling (buying) if the majority buys (sells) at time t+1. Thus, the dominant dynamics in this kind of market is that of a delayed majority game, in which agents benefit if their actions at time t are the majority actions at time t+1. Note that to a first approximation it does not matter whether an agent's action at time t is in the majority or minority at time t. All that matters is that the agent's action at time t is echoed by the majority at time t+1.[13]

This simple analysis indicates that in a neutral market no advantage is conferred on an agent's portfolio by "buying low" or "selling high". Portfolio gains can only be realized if subsequent market actions move the price in an advantageous direction. In a simplified model of the market, an agent's actions at time t will yield short term advantages only if the market moves the price in an appropriate direction. In the short term this implies the delayed majority dynamics described above. Of course, in a real market, different agents have different time horizons for their investment objectives. In the simple model we consider here, we are tacitly assuming that all agents have an investment horizon of one time step. The complications of heterogeneous investment horizons will be discussed elsewhere.

## IV. The Delayed Majority Game

These arguments indicate that delayed minority dynamics are important in determining the structure of at least a subset of real markets. To see what their implications are we will first study a very simple game, similar to the minority or majority games, but in which an agent receives a point (or a strategy improves its ranking) if and only if the group which an agent joins at time t (the group a strategy suggests joining at time t) turns out to be the majority group at time t+1. In this market interpretation, group 0 is the group of sellers and group 1 is the group of buyers. If the majority group at time t+1 is not the same as the group that the agent joined at time t, that agent does not receive a point. The architecture of the delayed majority game follows the pattern of the original minority game. Strategies are structured in the same way and decisions are based on which were the majority groups for the last m time steps. Note that in this delayed majority game there is no notion of price. That will be included in the somewhat more sophisticated market model in the next section.

As stated earlier, in the experiments we discuss below the publicly available information that the agents use to make their decisions is exogenous, so that a set of random numbers from 1 to $2^m$ are used as input to the strategy tables. The strategies that have successfully predicted the majority group one time step later are rewarded with 1 point, and at any moment, an agent uses his strategy that has the highest cumulative score. A coin flip is used to break ties when strategies are equally ranked.

The delayed majority games, both exogenous and endogenous have a relatively trivial structure similar to that of the endogenous majority game. Because of the way in which agents are rewarded, it will be beneficial to the majority of the agents if the majority group at time t is same as the majority group at t-1. In these games, therefore, after a transient period, most agents will be playing with a pure strategy and the system will approach an equilibrium state in which one group is chosen as the majority group most of the time. In a market interpretation, this corresponds to a perpetual buyer's (or a seller's) market. If we were to associate a price with the commodity in this interpretation, that price would either go to zero or diverge. (Note that this is different than the trivial behavior of the exogenous majority game in which there is not impetus for the majority group to be the same group from one time step to the next.)

## V.     The Market Game

Although delayed majority dynamics is clearly an important feature of a market, it is apparently not enough to generate interesting, let alone realistic structure. In particular, the behavior of the price (although it plays no explicit role in the dynamics of the delayed majority game) is an indication of pathology.

To rectify this situation, we note that one of the important features of any market is heterogeneity of beliefs. One aspect of this heterogeneity is embodied in the different strategies that agents use to make their choices about whether to buy or sell based on previous price movements. In the context of financial markets this can be viewed as heterogeneity in technical trading strategies. However, there is also heterogeneity in beliefs about the underlying value of a commodity that can be thought of as heterogeneity in fundamentals. To incorporate this heterogeneity, we let each agent have an innate expectation, $e_i$, of the intrinsic value of the commodity. If the current price is higher than an agent's expectation, and if the strategy that that agent is using suggests a "buy", then that agent will ignore the "buy" signal of its strategy, and will neither buy nor sell in that time step. Likewise, if the current price is lower than that agent's expectation, he will ignore a "sell" signal from the strategy he is playing. Thus, agents now have three possible actions. Agent i will join group 1 (buy) if his highest ranking strategy so indicates, and if $p(t)<e_i$ he will join group 0 (to sell) if his highest ranking strategy so indicates and if $p(t)>e_i$. Otherwise he will neither buy nor sell. This mechanism allows the number of agents participating in the market to vary over time. It also provides a natural mechanism for the system to correct run-away prices. In the simplest version of this game, which we shall describe here, the $e_i$'s are random values assigned to the agents at the beginning of the game and do not change with time. The delayed majority game, with the addition of intrinsic, heterogeneous price expectations, we call "the market game"[14].

To simplify our analysis, we choose the values of $e_i$ in a way such that log $e_i$ is uniformly distributed between two values, log $e_L$ and log $e_U$. $e_L$ and $e_U$ are selected such that they are symmetric about the initial price, i.e. $e_U/p(0)= p(0)/e_L$. We further define "expectation density", d, by $d \equiv N /(\log e_U - \log e_L)$. This is a measure of how densely the agents' expectations are packed into the price range ($e_L$, $e_U$). Note that the smaller d, the more relative heterogeneity there is in value expectation. Put another way, d controls the change in the number of agents who may choose not to trade (because of violated value expectations) given a certain price fluctuation.

To begin our analysis of the market game, we first consider the behavior of market *volatility* which is defined as σ[log p(t)], the standard deviation of the log of the price. Figure 3 shows market volatility as a function of m. Except for a set of small values of market volatility in the low-z region, there is an obvious qualitative similarity between this graph and Fig. 1 which shows $\sigma^2/N$ as a function of z for the standard minority game. To understand the relation between these figures, it is useful to first discuss some general aspects of price and portfolio behavior in the market game.

In the market game we associate a portfolio with each agent and a value to that portfolio. Initially we assign each agent an equal amount of cash and holdings. The value of the portfolio of each agent is calculated using the current commodity price. One interesting feature of these games is that the portfolios of agents tend to decrease in value over time. This can be attributed to the role of the market makers in these games, who make a profit thereby removing value from the system, which results in smaller portfolio values of the agents. Recall that $v_i(t) = m_i(t) + s_i(t)p(t)$, so $\overline{v(t)} = \overline{m(t)} + \overline{s(t)} \cdot p(t)$. When the market reaches equilibrium, the time-average of excess demand will be close to 0, so asymptotically $\bar{s}(t)$ would vary very slowly. Thus the long-term behavior of v(t) primarily depends on that of m(t). Consider the change in total cash holdings between two consecutive time steps:

$$\Delta m_{t,t+1} = -D(t) \cdot p(t+1) = -D(t)\exp(D(t)/\lambda)p(t) \tag{5}$$

Generally speaking, if D(t) is positive, p(t+1) is likely to be high due to a surfeit of buyers. This results in a fairly large negative Δm. On the other hand, if D(t) is negative, p(t+1) likely to be low, due to a surfeit of sellers, and so Δm will be positive but small. This suggests that the total cash holding of the agents will decrease over time. This can also be understood as the market makers taking advantage of the unbalanced supply and demand and exploiting the agents by buying low and selling high.

To prevent the game from halting, we allow the agents to continue to trade with negative cash and even negative portfolio values. Because the market makers remove value from the market, all agents will eventually have net negative portfolio values. (It is also easy to see that market makers remove value from the market at a higher rate when the volatility is high.) It may be unrealistic to expect highly indebted traders to continue to trade in a market. However, our purpose here is not to construct a highly realistic model of a market, but to exhibit how reasonable market dynamics can induce minority dynamics even when they are absent *ab initio*. Therefore, we ignore end effects such as agent bankruptcy.

Let us now return to a discussion of Fig. 3 and the relationship between the market game and the minority game. The position of the dip and the coalescence of values for different runs for larger values of m suggest a connection of these results with those of the original minority game, despite the fact that there is no *ab initio* minority dynamics in the market game. To better understand the relationship between the market game and the minority game, recall that in the market game, agents will not trade if the current price relative to their expectations is in conflict with the suggestion of their favored strategy. Consequently, an agent's actual trading behavior is not a direct reflection of the content of his favored strategy. It is possible, though, to observe another quantity that directly reflects the nature of agents' favored strategies, namely, the "potential number of buyers" $B_p(t)$. This is just the total number of the agents whose favored strategies indicate a "buy" (i.e. join group 1) whether or not the agent actually executes a trade. Similarly, we can also define the quantities "potential number of sellers" $S_p(t)$ and "potential excess demand" $D_p(t)$. The relationship between $D(t)$ and $D_p(t)$ will be discussed in more detail below, but first, consider Fig. 4. This shows the value of $\sigma^2/N$ of $B_p(t)$ as a function of m. Note the strong similarity between Figs. 3 and 4. In fact, we observe that within each game, there is a strong positive correlation between the standard deviation of $B_p(t)$ and market volatility. (Note also that this same correlation exists for $S_p(t)$ and $D_p(t)$, since they share the same standard deviation with $B_p(t)$). In addition, Fig. 4, which is just $\sigma^2/N$ for the potential membership of group 1, is very similar to Fig. 1. If we ignore the few small values in the low-m region (we shall discuss this below), we have a dip structure which is very similar to that in Fig. 1. Thus, there is an underlying dynamic in the market game which is that same as that of the minority game despite the fact that the original dynamics is that of a delayed majority game (supplemented by the dynamics of price and expectation). This dynamic is reflected not only in the standard deviation of the potential buyers (and sellers), but also, apparently, in the market volatility (Fig. 3). In this model of a market, minority dynamics are an emergent phenomenon.

To understand how minority dynamics are induced in the market game we first need to understand the behavior of successive price movements in the market game. To this end,, consider a binary series L which labels price movement. Let $L(t)=1$ if the price rises at time t (i.e., $p(t+1)>p(t)$) and let $L(t)= -1$ if the price falls ($p(t+1)<p(t)$). On the rare occasion that $p(t+1)=p(t)$, $L(t)$ is randomly chosen to be 1 or -1 with equal probability. Define $P(k)$ as an auto-correlation-function-like measurement of L in a game run for T time steps:

$$P(k) \equiv \frac{1}{T-k}\sum_{t=1}^{T-k}(1-L(t)L(t+k))/2 \qquad (6)$$

If there is no correlation between $L(t)$ and $L(t+k)$, $P(k)$ is expected to be 0.5. Smaller (larger) values indicate that $L(t)$ and $L(t+k)$ are more likely to be the same (different). Of particular interest is the value of $P(1)$, which deals with the relationship of price movements in consecutive time steps. We will show below that the value of $P(1)$ is directly related to the extent to which minority dynamics is present in the market game. We note, first of all, that in the market game, $P(1)$'s tend to be greater than 0.5, so that there is a tendency for price movements at adjacent time steps to be different. This can be qualitatively understood as follows: Suppose, for example, that $p(t+1)>p(t)$. Consider the group of agents that have price expectation values between $p(t)$ and $p(t+1)$. Call this group, group A. At time t, these agents will not sell even if their strategies say "sell", and at time t+1, they will not buy. Denote the group of buyers at time t by $B(t)$ and the group of sellers by S(t). On average, about half of the members of group A belong to B(t) and the other half belong to $S_p(t)$ (i.e. their favored strategy indicates a "sell", but there price expectation indicates that the current price is too low). The selling and buying populations outside A should remain approximately remain the same size from time t to t+1, since the relative size of their price expectations compared the actual price does not change from t to t+1. Therefore, $|B(t+1)| \approx |B(t)| - |A|/2$ and $|S(t+1)| \approx |S(t)| + |A|/2$ (|X| denotes the size of group X). As a result, $D(t+1) = |B(t+1)| - |S(t+1)| \approx D(t) - |A|$. So, if D(t) is not too large, the price movement at time t+1 is more likely to be negative (due to a surfeit of sellers), opposite to that of time t. A similar argument applies if p(t+1)<p(t), and so P(1) will generally be greater than 0.5 in the market game.

Using now the definition of the expectation density, *d* (and recalling that we assume that the agents are uniformly distributed in log of price expectation) we have: $|A| \approx (\log(p(t+1)) - \log(p(t))) \cdot d$, and so

$$\begin{aligned} D(t+1) &\approx D(t) - (\log p(t+1) - \log p(t)) \cdot d \\ &= D(t) - (\log p(t+1) - \log p(t))\lambda \cdot (d/\lambda) \\ &= D(t)(1 - d/\lambda) \end{aligned} \qquad (7)$$

Returning now to P(1), we note from (6) that P(1) is determined by the time series of the sign of D(t). Although, in principle, this complete time series is necessary to exactly determine P(1), we have found that the most important dependence of P(1) is on d/λ. In fact, our results are consistent with a scaling behavior for P(1), in which there is no residual dependence on d (or N) separately once d/λ is specified. (In addition, as we shall show below, this same scaling curve obtains for all values of m greater than a critical value.) In fig. 5 we plot P(1) as a function of d/λ for m=8 and for a range of different values of N and λ. P(1)'s of various λ and d lie in the same curve (an estimated expression for the linear portion is $P(1) = 0.25 d/\lambda + 0.5$). For any positive value of d/λ, there is a negative correlation between L(t) and L(t+1) (P(1) >0.5). For d/λ greater than 2, the negative correlation becomes complete and P(1) saturates at 1. Note that this is consistent with the approximation, (7), which, for d/λ=1 implies that $D(t+1) = -D(t)$.

With this background we can understand how minority dynamics are induced in the market game. By construction, delayed majority dynamics rewards strategies that instructed (or would have instructed) an action at time t which turns out to be the majority action at time t+1. But we have seen that in the market game there is a strong anti-correlation between consecutive price movements, (as indicated by the fact that P(1)>0.5) which means that a majority action at time t+1 is a *minority* action at time t. So, for example, in the case discussed above, the majority action at time *t+1* is *most likely* to be "sell". Therefore, a strategy that said "sell" at time t will be rewarded. But since p(t+1)>p(t), selling is the *minority* action at time t.

This argument captures the essence of the dynamics that induces minority behavior. However, it is incomplete. In the argument of the last paragraph, the minority groups refer to the groups of actual-buyers and actual-sellers that are smaller. We can also ask whether the induced minority dynamics applies to the groups of potential-buyers and potential-sellers. It is true that an analysis that focuses on the groups of potential buyers and sellers is further removed from the actual outcome of the market game in terms of wealth accrual than is the analysis that emphasizes the groups of actual buyers and sellers. However, analyzing the game with respect to the groups of potential buyers and sellers focuses our attention on the underlying dynamics of strategy choice of the agents, which is at the heart of the fundamental dynamics of the minority game, and is therefore a worthwhile enterprise.

In fact, to really understand whether there are underlying minority dynamics, we will need to study cross-correlation functions, that relates the sign of D(t) with the sign of $D_p(t)$. The reason is that it is the sets of potential buyers and sellers that carry the information about minority dynamics (which reflects which of an agent's strategies are dominant), but it is outcome of the *actual* trades that determine the rewards to those strategies, and that is determined by the sets of *real* buyers and sellers.

To that end, let G(t) be the sign of $D_p(t)$, which carries the information of which was the minority group between potential buyers and sellers at time t. I.e. G(t)=1 if $B_p(t)>S_p(t)$ and G(t)=-1 otherwise. For the emergent dynamics to be that of a minority game, we require that the minority group of *potentials* at time t-1 is the same as the majority group of *actuals* at time t. A measure of how often this happens is the cross-correlation function $\rho_1$ defined as:

$$\rho_1 = \frac{1}{T-1}\sum_{t=2}^{T}(1-G(t-1)L(t))/2 \tag{8}$$

Of course, a measure of the extent to which the game is a majority game in terms of the underlying potential buyers and sellers is just $1-\rho_1$.

Another quantity of interest measures the extent to which the underlying groups of potential buyers and sellers plays a delayed-majority game. (This is not to be confused with the *ab initio* delayed-majority game which defines our market system.) The way in which our market game is constructed dictates that strategies that predict, at time t-1, the actual majority group at time t will gain a point. But we want to know the extent to which the market game instantiates dynamics of a delayed-majority game only with repect to the groups of potential buyers and sellers. Therefore, we need to know what the correlation is between the actual and potential majority groups at time t. If that correlation is complete, then the market game is also a delay-majority game in terms of the potential buyers and sellers. We thus define $\rho_2$ as

$$\rho_2 = \frac{1}{T}\sum_{t=1}^{T}(1+G(t)L(t))/2 \tag{9}$$

the larger $\rho_2$ the more likely it is that the majority group of *actuals* (buyers or sellers) will be the same as the majority group of *potentials* (buyers or sellers) at any given time, and so the more likely it will be that the emergent dynamics will be that of a delayed-majority game among the groups of potential buyers and sellers.

We want to describe the ways in which the $\rho_I$'s vary among games with different parameter settings. We will argue that, like P(1) the $\rho_I$'s largely scale among different games with d/l. To do that, we first discuss the relationship between D(t) and $D_p(t)$. We make the reasonable assumption that in any range of log(price) potential buyers and potential sellers are both uniformly distributed, and the probability of a given agent being a potential buyer or a potential seller is 0.5. With this assumption, a brief calculation show that

$$D(t) = \frac{1}{2} D_p(t) + (\log \bar{p} - \log p(t))d \qquad (10)$$

where $\log \bar{p} = (\log e_U - \log e_L)/2$. If we replace t with t+1, take the difference between the two equations, and, using (1), we have

$$D(t+1) - D(t) = (D_p(t+1) - D_p(t))/2 - (\log p(t+1) - \log p(t))d =$$
$$(D_p(t+1) - D_p(t))/2 - D(t)d/\lambda$$

so

$$D(t+1) - D(t)(1 - d/\lambda) = (D_p(t+1) - D_p(t))/2 \qquad (11)$$

Note that if we assume that $D_p(t)$ is time independent, we recover the approximation (7). We have seen that P(1) is a function only of $d/\lambda$ (Fig. 5). From (11), it appears that the only other variable that is involved in the relation between D(t) and $D_p(t)$ is $d/\lambda$. Therefore, it is reasonable to assume that functions that correlate the signs of D(t) with $D_p(t)$ (i.e., the $\rho_i$'s will be a function only of $d/\lambda$.

Fig. 6 shows $\rho_1$ and $\rho_2$ as a function of $d/\lambda$ for m=8 and different values of N. (The behavior for smaller m will be discussed below.) P(1) is also plotted for purposes of comparison. From this figure we can deduce the extent to which minority dynamics and delayed majority dynamics among groups of potential buyers and sellers emerges in different games. There are 3 regions in this graph: for $0<d/\lambda<1$, $\rho_2$ is larger than $\rho_1$. For $1<d/\lambda<2$, $\rho_1$ is larger than $\rho_2$. When $2<d/\lambda$, both $\rho_2$ and $\rho_1$ are near 0.5. When $d/\lambda<2$, delayed majority dynamics among the potentials is always present since $\rho_2 >0.5$. When $d/\lambda>2$, there is neither delayed majority dynamics nor minority dynamics. Recall equation (7), D(t+1)=D(t)(1- d/$\lambda$). When d/$\lambda$>2, not only will D(t+1) has different sign with D(t), its magnitude is also greater than that of D(t). This implies an amplification of the magnitude of D(t) as a function of time. |D(t)| will until it reaches a limit set by the number of traders. In principle, the natural limit of D(t) is N, the total number of the traders. In fact, though, the maximum D(t) is usually about N/2 due to the nearly equal number of potential sellers and buyers, and this limit occurs when price is greater than $e_U$ or less than $e_L$. Due to the alternating signs of D(t) (and its large magnitude), the price alternates between very high and very low values. As a result, P(1) is always 1, and $\rho_1$ and $\rho_2$ are always close to 0.5 because the potential minority groups have no correlation with the actual minority groups, which are now just switching back and forth between buyers and sellers.

When $\rho_1>0.5$ the market game manifests underlying minority dynamics among the groups of potential buyers and sellers. This occurs for all values of $d/\lambda<2$, but interestingly, the extent to which there is such minority dynamics peaks when $d/\lambda$ is near 1.3. To understand this, note that the criteria for the presence of minority dynamics is that the strategy that would have put an agent in a minority group is being rewarded. In the market game, the strategy that predicts the *actual* majority action at the next time step is rewarded. Therefore, for the minority dynamics to exist for the underlying potential groups, the *actual* majority group at time t must be the same as the *potential* minority group at time t-1. For this to happen, it must be that either 1) the actual majority group at time t-1 is *opposite* to the actual majority group at time t, and the actual majority group at time t-1 is the *same* as the potential majority group at time t-1, or, 2) the actual majority group at time t-1 is the *same* as the actual majority group at time t, and the actual majority group at time t-1 is *opposite* to the potential majority group at time t-1. This implies a relationship between P(1), $\rho_1$ and $\rho_2$. Recall that P(1) is the probability that the actual majority group at time t-1 is the *opposite* to the actual majority group at time t, and $\rho_2$ is the probability that actual majority group at time t is the same as the potential majority group at time t. When $d/\lambda$ is small, P(1) is near 0.5 and $\rho_2$ is close to 1. Thus, although the potential groups and the actual groups are almost always the same at a given time, there is a probability of about 0.5 that the actual majority group at time t-1 is *opposite* to the actual majority group at time t. As a result, condition 2) will never be satisfied and condition 1) will be satisfied about 50% of the time. Therefore, minority dynamics will not dominant when $d/\lambda$ is very small. A similar argument obtains in the case in which $d/\lambda$ is greater than 2. Here P(1) is about 1 and $\rho_2$ is close to 0.5, so it is easy to see that condition 1) will never be satisfied, and condition 2) will be satisfied about 50% of the time. It is only when $d/\lambda$ is of moderate value and both P(1) and $\rho_2$ are greater than 0.5, that minority dynamics among the groups of potentials can be dominant ($\rho_1>0.5$).

Although the actual dynamics is more complex, a simple qualitative explanation of the shape of the $\rho_1$ curve can be obtained if we assume independence between P(1) and $\rho_2$. Under this assumption, it is easy to show that

$$\rho_1 \approx P(1)\rho_2 + (1-P(1))(1-\rho_2) \tag{12}$$

The curve (12), using the empirically determined values of P(1)'s and $\rho_2$'s, is plotted as the dotted line in fig.6. This curve has the correct shape of the empirical curve for $\rho_1$, and peaks at the correct value of $d/\lambda=1.3$, where P(1)= $\rho_2$.

The difference between the approximation (12) and the empirical values of $\rho_1$ is that the empirical results are consistently larger than those indicated by (12). This difference is consistent with a positive correlation between P(1) and $\rho_2$. That is, whenever the actual majority group at t-1 is the same as the potential majority group at time t-1, there is an enhanced probability that the actual majority group at time t-1 and the actual majority group at time t will be opposite. This positive correlation can be understood by considering, for example, the case in which at time t-1 there is a majority of potential buyers. If the price at time t-1 is not too high, there will be a high probability for a majority of actual buyers at time t-1. This will tend to force the price up at time t, and so will decrease the chance that the group of actual buyers will be a majority at time t. That is, it will increase the chance that the actual majority group at time t will be sellers. In this scenario, the group of potential sellers at time t-1 (who form the minority group of potentials at t-1 in this example) will be rewarded. This is exactly the minority dynamics. Thus, in the market game, induced minority dynamics predominates when $0<d/\lambda<2$, and is particularly pronounced when $1<d/\lambda<1.5$. At its peak, when $d/\lambda\approx1.3$, the market game is about 78 percent minority dynamics. This minority dynamics is due both to the intrinsic structure of the market game, and to the induced positive correlation between P(1) and $\rho_2$.

The presence of induced minority dynamics helps explain the behavior of $\sigma^2/N$ of the potential buyers and the price volatility in the high m region. Consider, for example, N=101, $\lambda$=100 and two values of d, d=150 and d=190. When N=101 and $d/\lambda$=1.5 ($d/\lambda$=1.9), for all m>6, $\rho_1$ is about 0.8 (0.55) implying that induced minority dynamics dominates at about the 80% (55%) level. In Fig. 7 we compare $\sigma^2/N$ of potential buyers in these market games with that of the simple mixed minority-majority game described in section II, with a matched percentage of minority dynamics. The similarity is obvious, even though the minority dynamics in the market game is induced, while the minority dynamics in the simple mixed game is inserted *ab initio*. Comparisons for price volatility between the mixed majority-minority game and the market game are similar.

These arguments are only valid when m is greater than 6. (This number may be slightly N dependent, but see eq. (13) for a semi-quantitative derivation of this cross-over value of m.) For m below this value, different dynamics apply. Fig. 8 shows P(1), $\rho_1$ and $\rho_2$ as function of $d/\lambda$ for m=2. It can be seen that, when $d/\lambda<2$, neither delayed majority dynamics nor minority dynamics are dominant since both $\rho1$ and $\rho2$ are close to 0.5. Behaviors in this parameter range will be discussed in next paragraph. When $d/\lambda>2$, all of the three quantities, P(1), $\rho_1$ and $\rho_2$, reach an approximately constant value. These values are also m-dependent. The value of $\rho_1$ in this region is greater than 0.5, which would seem to imply that the game played here is a mixed minority-majority game with some fraction of minority dynamics. However, the dynamics here are very different from the simple mixed minority-majority game. First note that the step-function-like behavior at $d/\lambda=2$ is caused by the generation of large price fluctuations for $d/\lambda>2$, as described above for the high m case. However, when m is small, another dynamic is at play which changes the behavior of the system and lead to different values of P(1) and the $\rho_i$'s. We will first explain the behavior of the system for low m and $d/\lambda<2$, following which we will return to discuss the behavior for low m and $d/\lambda>2$. Some of the results in the former will be helpful in our discussion of the latter.

We now provide a qualitative description on the behaviors of the games in the low-m phase with $d/\lambda<2$. In these games, there is a biased state in which one trading action is strongly favored over the other. In "regular" market games (or states), we expect that the long-term average of the commodity price is close to the median of the distribution of price expectations of the agents. If the initial price p(0) is close to this value, then the long-term average price will also be about the same as the initial price p(0). As is usual with the laws of supply and demand, if the price significantly deviates from the median of expectations say, when it is much higher, then there will be more agents actually selling than buying and the excess supply will lower the price dragging it back to "normal". This is the general picture of the price movement in a market game, assuming that on average each agent has equal probability to sell or to buy, which means that at a given time, the number of potential traders is not biased.

In a "biased state", this assumption doesn't hold. For example, in a buying-biased state, most agents are stuck with a strategy that favors buying over selling. (I.e. potential buyers greatly outnumber potential sellers). In such a state a low to medium price will be driven up by a surplus of buyers. But because potential buyers outnumber potential sellers, the equilibrium price to which the commodity is driven, $p_{eq}$, will not be close to the median of the agents' expectations. When the price is near $p_{eq}$ there is nearly equal probability for the price to rise or fall, but the potential majority group of seller or buyers (in this case, buyers) is always the same, so $\rho_1$ and $\rho_2$ are close to 0.5. P(1) still increases with increasing $d/\lambda$, due to the increased negative correlation between D(t) and D(t+1).

This kind of biased states can be easily reached, particularly when m is low. Recall that the fundamental *ab initio* dynamics in our market game is delayed majority dynamics, so there is a natural tendency to reach a steady state that favors one action. Consider the following scenario, which can easily occur, particularly with smaller m: a significant majority of agents' higher-ranking strategies have more 1's (buy) than 0's (sell), (This situation is easier to realize for small m, because there is less heterogeneity among agents' strategies when m is small.) For most input signals, this configuration will produce a response of buyer-majority. If, for several consecutive time steps, the randomly-generated input signals are such that the responses are all buyer-majority, the price will go up, reinforcing the buying bias of most agents, which will lead to further price rises.

Such a state is maintained by moderate separations between the ranks of each agent's two strategies. Since both $\rho_1$ and $\rho_2$ are near 0.5, there is no *a priori* tendency for the rank differences between the strategies to decrease, leading to a reasonable persistence of the biased state. Generally, the biased state will be broken by random fluctuations, after which there will be a transient period in which no trading action is preferred. However, in the low-m region the system will soon go back to one or the other biased state

It is possible to derive a relationship between $p_{eq}$ and m which can be used to explain the value of the cross-over from low-m to high-m dynamics in our game. The equilibrium price is directly related to the fraction of the agents choosing a particular action when in a biased state. For example, the equilibrium price in a buyer-biased state is given by: $\log(p_{eq})=\log(e_L)+(\log(e_U)-\log(e_L))d_b$, where $d_b$ is the ratio of (potential buyers)/N. On average, $d_b$ is given by the formula

$$d_b = \frac{1}{2^m}\sum_{l=0}^{2^m} l \left( \sum_{i=0}^{l-1} 2 \binom{2^m}{l}\binom{2^m}{i} / 2^{2^{m+1}} + \binom{2^m}{l}\binom{2^m}{l} / 2^{2^{m+1}} \right) \tag{13}$$

$d_b$ is a rapidly decreasing function of m. When m > 5, $d_b$ is very close to 0.5, which means that the equilibrium price is almost the same as the median of agents' expectations. Thus, for m>5, it is difficult for the system to enter and maintain a biased state.

The effect of the biased states on the price volatility is primarily manifested through the switching between the buying-biased-states and selling-biased-states. During the epoch of one biased state, the price volatility is small because of the reduced number of actual traders, which explains those data points with small volatility observed in figure 3. But if the time period of interest spans opposite biased states, the volatility will be high. Indeed, if we observe the runs shown in Fig. 3 for a longer time, the low volatility results for small m will rise. In addition, since in the low-m regions, lower m leads to larger separation of the equilibrium prices of the two opposite (buying and selling)-biased states, price volatility (upper branch) increases with decreasing m.

We now return to a brief explanation of the case $d/\lambda>2$ and small m. Recall that in the case of $d/\lambda>2$ and large m, the price alternates with each time step between very high and very low values. For low m, there is a high probability that an extremely high (low) price is accompanied by a state in which most agents belong to the group of potential buyers (sellers). This causes the magnitude of D(t+1) to be small so that the price at time t+1 reverts to lie between $e_L$ and $e_U$. This increases the probability that D(t+2) is *not* of opposite sign of D(t+1) thus effectively reducing P(1). The increase in $\rho_1$ from 0.5 can be explained in a similar way.

Although of some passing interest, the case in which $d/\lambda>2$ is not of central importance. In our model agents' expectations are fixed from the beginning of the game and so $e_L$ and $e_U$ are fixed parameters. But in real markets, the expectations of agents can (and do) change with their observations of price history. So, if the price gets very large, agents' expectations will change, and the price will still fluctuate between (newly established values of) $e_L$ and $e_U$. The dynamics of wild price fluctuations outside of the range of expectations will generally not occur, rendering our model with $d/\lambda>2$ unrealistic.

To summarize, in figure 9 we have plotted a qualitative phase diagram for our market game, identifying different parameter regions of $d/\lambda$ and m associated with different behaviors. Of the four phases represented in the figure, one of them gives rise to the induced minority dynamics (high m and $d/\lambda<2$) and other phases are dominated by other (generally less interesting) behaviors. As a function of $d/\lambda$, there is a phase change exactly at 2. As a function of m, there is a phase change for m near 5 or 6.

## VI. Discussion and Conclusion

### A. Relationship to Another Market Model

Before presenting our conclusions, it is worthwhile to briefly compare our approach to another model of markets that explicitly considers minority dynamics. We do this for two reasons. First, because there are quite fundamental differences in the two approaches, despite superficial similarities. Second, the differences emphasize interesting questions about what ought to be included *ab initio* in models of financial markets. The model we have in mind is that discussed by Marsili[1], which incorporates delayed dynamics and heterogeneous agent beliefs. Although the explanation in this paper is somewhat different, the basic idea is that there are two types of agents, fundamentalists and trend-followers. The former presume that winning trades are those that place them in the minority of buyers or sellers at a given time step, while the latter, are traders who presume that winning trades are those that place them in the majority of buyers or sellers at a given time step. Changes in portfolio values are not directly relevant in determining the efficacy of strategies according to each agent. Rather, an agent's strategy is most valued if its actions accord with the expectations of that agent. Marsili discusses the extent to which the resulting game is a majority or minority game, and the extent to which price behavior is consistent with agents' expectations.

The approach in our work is quite different, in that the heterogeneity of our agents is not based on different expectations about the dynamics of the market (cf. fundamentalist or trend-followers), but on heterogeneous beliefs about the underlying value of the commodity being traded. Furthermore, the success of an agent's strategies in our approach is evaluated directly on the basis of whether that strategies actions led to an increase or decrease in the portfolio value of the agent. (In this regard, the difference between Ref. 1 and our work parallels the difference pointed out earlier in mixed majority-minority games between "mixed-populations" and "temporal-mixing".) In the work of Ref. 1, minority dynamics and majority dynamics is inserted into the model *ab initio* in the form of heterogeneous beliefs by the agents about market dynamics. In our model, minority dynamics emerges from the interaction of agents whose fundamental heterogeneity is associated with different beliefs about the commodity's underlying value.

Real financial markets are extraordinarily complex, in no small part due to the complexity of the agents (traders). The work in Ref. 1 highlights and models one source of heterogeneity in traders. Our work highlights and models another. To what extent these sources of heterogeneity are important in different markets at different times is a complex question, well beyond the scope of this discussion. However, the study of a range of models that isolate different aspects of heterogeneity is clearly useful in trying to isolate and understand the consequences of different kinds of heterogeneous beliefs.[15]

### B. Summary and Conclusion

Very simple arguments suggest that minority dynamics express a fundamental dynamic of financial markets in the sense that "buying low" or "selling high" can be understood as minority actions in a market driven by a very simple supply-demand relation. However, these arguments are inadequate in that they do not take into account realistic mechanisms by which agents determine value based on predictions of price movements. In this paper, we argued that delayed majority dynamics is a simple version of a realistic and primary dynamic in many financial markets. We then showed that a simple delayed majority game is trivial and has uninteresting behaviors. However, when a game with delayed majority dynamics is modified to include heterogeneous price expectations for the agents (which allows agents to choose not to trade if their price expectations and preferences for action are inconsistent), the emergent behavior is much more interesting. In particular, we find that for a reasonable range of parameter settings, (see below) this game (the Market Game) exhibits emergent dynamics consistent with that seen in the Minority Game, despite the fact that minority dynamics are not introduced *ab initio*. We furthermore showed that the introduction of heterogeneous price expectations has the happy consequence of lowering price volatility. The important controlling variable in the Market Game is the ratio between the density of expectations, d, and the liquidity, $\lambda$. When $0<d/\lambda<2$ and $m>6$, minority dynamics are induced and the market game exhibits minority game behavior. Finally, we discussed the general phase structure of the Market Game as a function of the control parameters, $d/\lambda$ and m. For other values of these parameters qualitatively different (and in general less interesting) behaviors were observed for which we provided qualitative explanations.

It is most gratifying that minority dynamics emerges in the Market Game. This result justifies our most basic intuition about the role of supply and demand in determining value, and shows how the primacy of minority behavior ("buy-low, sell-high") emerges from a model that embodies more realistic fundamental dynamics.

We are very grateful to Michael Wellman for many helpful suggestions and stimulating discussions on this topic.

# Figure Captions

1. $\sigma^2/N$ vs. Z for minority game with N=101, 16 runs for each parameter setting
2. $\sigma^2/N$ vs. Z for mixed minority-majority game. N=101. Each line represents set of games with certain *p*. From bottom to top, *p* ranges from 0% to 100%, in 10% increments
3. Market volatility of market game with N=101. Volatilities are calculated on the last 2,000 time steps of simulations of 10,000 time steps.
4. $\sigma^2/N$ of the number of "potential buyers" in the market game in figure 3.
5. P(1) as a function of d/$\lambda$ for market games with different *d* and $\lambda$ settings.
6. ρ1 and ρ2 as a function of *d/$\lambda$* for market games with different *d* and $\lambda$ settings, m=8.
7. $\sigma^2/N$ of the potential buyers of market game and the $\sigma^2/N$ of the mixed minority/majority game with the same proportion of the minority dynamics.
8. $\rho_1$, $\rho_2$ and P(1) as function of d/$\lambda$ for market games with different *d* and $\lambda$ settings, m=2.
9. Qualitative phase diagram of the market game.